\documentclass[a4paper, aps, amsmath, 10pt, nofootinbib]{revtex4-2}

\usepackage{tikz}
\usepackage{physics,caption,braket}
\usepackage{xcolor}
\usepackage{amsfonts}
\usepackage{amsmath}
\usepackage{hyperref}
\usepackage[normalem]{ulem}
\usepackage[capitalize]{cleveref}

\hypersetup{hidelinks}
\usetikzlibrary{arrows.meta}

\newcommand{\braopket}[3]{\left\langle{#1}|{#2}|{#3}\right\rangle}

\begin{document}

\title{On Quasiparticles within the Refined Gribov-Zwanziger Model}

\author{Felipe F. Garcia}
\email{felipefernandesgarcia96@gmail.com}
\affiliation{UERJ -- Universidade do Estado do Rio de Janeiro,	Instituto de Física -- Departamento de Física Teórica -- Rua São Francisco Xavier 524, 20550-013, Maracanã, Rio de Janeiro, Brazil}

\author{Marcio A. L. Capri}
\email{caprimarcio@gmail.com}
\affiliation{UERJ -- Universidade do Estado do Rio de Janeiro,	Instituto de Física -- Departamento de Física Teórica -- Rua São Francisco Xavier 524, 20550-013, Maracanã, Rio de Janeiro, Brazil}

\author{Bruno W.~Mintz}
\email{brunomintz@gmail.com}
\affiliation{UERJ -- Universidade do Estado do Rio de Janeiro,	Instituto de Física -- Departamento de Física Teórica -- Rua São Francisco Xavier 524, 20550-013, Maracanã, Rio de Janeiro, Brazil}

\begin{abstract}
The problem of the effective excitations of a gauge theory in the nonperturbative regime remains not completely understood. Using the Refined Gribov-Zwanziger (RGZ) theory as an effective model for pure gauge Yang-Mills theories, we revisit the quasiparticle (quadratic) excitations of the theory. A novel interpretation of such quasiparticles is proposed, in the case of real mass poles, taking into account the manifest ${\cal PT}-$symmetry of the RGZ Lagrangian. \end{abstract}

\maketitle
\date{}

% --------------------------------------------------
\section{Introduction}
% --------------------------------------------------

Non-Abelian gauge theories provide the fundamental description of the strong interaction, yet their infrared regime remains one of the most challenging sectors of quantum field theory \cite{Peskin:1995ev, bib:ryder}. In particular, the confinement of gluons and quarks and the corresponding analytic structure of gauge-dependent correlation functions continue to motivate the search for effective descriptions capable of capturing the relevant nonperturbative features of Yang--Mills theory. Among the conceptual obstacles in this direction lies the problem of gauge fixing beyond perturbation theory.

The standard Faddeev--Popov procedure successfully implements gauge fixing in the perturbative regime \cite{Faddeev:1967fc}. However, as first pointed out by Gribov \cite{Gribov_1978}, it does not completely remove gauge redundancy in non-Abelian gauge theories. In the Landau gauge, multiple gauge-equivalent field configurations may still satisfy the same gauge condition, giving rise to the so-called Gribov copies. In order to deal with this problem, Gribov proposed restricting the domain of integration in the functional integral to the so-called Gribov region. Zwanziger later reformulated this restriction through a nonlocal horizon term and its local representation in terms of auxiliary bosonic $(\bar\varphi^{ab}_{\mu},\varphi^{ab}_{\mu})$ and fermionic $(\bar\omega^{ab}_{\mu},\omega^{ab}_{\mu})$ fields, thereby giving rise to the Gribov--Zwanziger framework \cite{Zwanziger_1989}.

Subsequent developments showed that the condensates of mass dimension $d_m=2$ given by $\langle A_\mu^aA_\mu^a\rangle$ and $\langle\bar\varphi_\mu^{ab}\varphi_\mu^{ab}-\bar\omega_\mu^{ab}\omega^{ab}_\mu\rangle$ are dynamically generated \cite{Sorella_a_refinement,DudalSorellaVandersickel2011}. In the Refined Gribov--Zwanziger (RGZ) model, effective mass parameters emerge both in the gluon sector and in the localizing-field sector, improving the infrared behavior of the theory and leading to propagators compatible with positivity violation \cite{Vandersickel:2012tz} and lattice-inspired results \cite{DudalOliveiraSilva2018}. Similar conclusions have been found by other nonperturbative methods, such as the Dyson-Schwinger equations \cite{AguilarBinosiPapavassiliou2008} and the Functional Renormalization Group \cite{FischerMaasPawlowski2009}. In this framework, the gluon propagator displays a pair of complex-conjugate poles, often called \emph{i-particles} \cite{Sorella_i_particles,DudalOliveiraSilva2018}. These field excitations cannot correspond to physical asymptotic states, since their propagators do not possess a well-defined K\"allen-Lehmann representation \cite{Kallen:1952zz,Lehmann:1954xi}. Such property is often interpreted as a consequence of color confinement \cite{Alkofer:2000wg,Cucchieri_2005,2013MPLA...2830035C}. In a context of complex mass poles, the \emph{i-particles} can be considered an analogue of quasiparticles in such theories \cite{Sorella_i_particles,Mintz:2024soo}.

In the present work, we explore in detail how the color and Lorentz structure of the fields of the theory define ``effective quasiparticle'' fields at the quadratic level of the RGZ Lagrangian. As a new step towards a consistent interpretation of such fields, we first discuss the pseudohermiticity of the RGZ action, following the ideas of Bender, Mostafazadeh and others \cite{Bender_1998,Mostafazadeh_2010_07}. Then, for the case of real mass poles, we observe that the quasiparticle field operators can be made Hermitian with respect to a positive inner product. (Of course, the gauge dependence of such fields makes them not physical observables in a strict sense.) This result is obtained by showing that the propagator of the quasiparticles has a well-defined, positive spectral function, as long as the mass poles $p^2=-M_\pm^2<0$ are real\footnote{We work in Euclidean space throughout this paper.}. The case of complex mass poles is also very interesting (due to its known connection with the gluon propagator \cite{DudalOliveiraSilva2018}), but is much more subtle and will be studied in a future work. Based on this picture, we discuss how the degrees of freedom of the theory could be better described by the quasiparticles than by the fundamental fields, such as $A_\mu^a$, in a high energy regime. 

This paper is organized as follows. After a brief review of the local RGZ action in Sec. \ref{sec:RGZreview} we derive the quasiparticles from the quadratic action in Sec. \ref{sec:iparticles}. In section \ref{sec:pseudohermiticity}, we discuss the pseudohermiticity of the RGZ action and use this property to interpret the quasiparticle fields in the limit of real mass poles. Finally, Sec. \ref{sec:conclusions} summarizes our conclusions and perspectives for future studies.

% --------------------------------------------------
\section{The Refined Gribov--Zwanziger model in the Landau gauge}
\label{sec:RGZreview}
% --------------------------------------------------

We begin by recalling the local formulation of the Refined Gribov--Zwanziger model in the Landau gauge \cite{Zwanziger_1989,Sorella_a_refinement}. In Euclidean space and in Landau gauge, the RGZ action can be written as
\begin{equation}\label{eq:S_RGZ_def}
S_{RGZ}=S_{FP}+S_H+S_{\text{cond}},
\end{equation}
where the Faddeev--Popov sector is given by
\begin{equation}
S_{FP}
=
\int d^d x
\left(
\frac{1}{4}F_{\mu\nu}^aF_{\mu\nu}^a
+ i b^a \partial_\mu A_\mu^a
-\bar c^a \mathcal{M}^{ab}(A)c^b
\right),
\end{equation}
with the Faddeev-Popov operator in the Landau gauge given by
\begin{equation}
\mathcal{M}^{ab}(A)
=
-\partial_\mu D_\mu^{ab}(A)
=
-\delta^{ab}\partial^2
+
g f^{abc}A_\mu^c\partial_\mu,
\end{equation}
since the covariant derivative reads $D_\mu^{ab}:=\delta^{ab}\partial_\mu-gf^{abc}A_\mu^c$.
The local horizon sector is
\begin{equation}\label{eq:local-horizon}
S_H =
\int d^d x
\left[
\bar\varphi_\mu^{ac}\mathcal{M}^{ab}(A)\varphi_\mu^{bc}
-
\bar\omega_\mu^{ac}\mathcal{M}^{ab}(A)\omega_\mu^{bc}
+
i g\gamma^2 f^{abc}A_\mu^a\left(\bar\varphi_\mu^{bc}+\varphi_\mu^{bc}\right)
-d(N_c^2-1)\gamma^4
\right].
\end{equation}
The action $S_{GZ}=S_{FP}+S_{H}$ is known as the Landau gauge local Gribov-Zwanziger (GZ) action \cite{Zwanziger_1989}. It can be shown that it is unstable with respect to the formation of mass dimension $d_m=2$ condensates $\langle A_\mu^a(x)A_\mu^a(x)\rangle\not=0$ and $\langle\bar\varphi_\mu^{ab}(x)\varphi_\mu^{ab}(x)-\bar\omega_\mu^{ab}(x)\omega_\mu^{ab}(x)\rangle\not=0$. In order to reinforce such nonzero condensates directly in the starting action, one usually directly adds the term 
\begin{equation}
S_{\text{cond}}
=
\int d^d x
\left[
\frac{m^2}{2}A_\mu^aA_\mu^a
+
M^2\left(
\bar\varphi_\mu^{ab}\varphi_\mu^{ab}
-
\bar\omega_\mu^{ab}\omega_\mu^{ab}
\right)
\right],
\end{equation}
so that the parameters $m^2$ and $M^2$ can be regarded as Lagrange multipliers that reinforce the presence of nonzero condensates. The resulting action (\ref{eq:S_RGZ_def}) is then called the Refined Gribov-Zwanziger (RGZ) action. 
Following the Lagrange multiplier method, the nonzero Gribov parameter $\gamma$ \cite{Gribov_1978} can be obtained as a solution to the gap equation
\begin{eqnarray}
    \left.\frac{\partial E_{vac}}{\partial\gamma}\right|_{\gamma\not=0}=0,
\end{eqnarray}
where $E_{vac}$ is the vacuum energy. However, in many studies, the Gribov parameter, as well as $m^2$ and $M^2$, are left as adjustable parameters\footnote{In self-consistent theory, $m^2$, $M^2$, and $\gamma^2$ would all be obtained self-consistently from their respective (and coupled) gap equations.}.

It is convenient to combine the mass term in the localizing sector with the Faddeev--Popov operator by introducing the massive operator
\begin{equation}
\mathcal{M}_M^{ab}(A)
=
-\partial_\mu D^{ab}_\mu(A)+M^2\delta^{ab}.
\end{equation}
 We also denote the standard Faddeev-Popov operator as ${\cal M}\equiv[{\cal M}]_{M=0}$. In this notation, the classical RGZ action reads

\begin{equation}\label{eq:RGZ-action-full}
\begin{aligned}
S_{RGZ}
=
\int d^d x \Bigg[
&\frac{1}{4}F_{\mu\nu}^aF_{\mu\nu}^a
+i b^a\partial_\mu A_\mu^a
-\bar c^a\mathcal{M}^{ab}(A)c^b \\
&+
\bar\varphi_\mu^{ac}\mathcal{M}_M^{ab}(A)\varphi_\mu^{bc}
-
\bar\omega_\mu^{ac}\mathcal{M}_M^{ab}(A)\omega_\mu^{bc}
+
i g\gamma^2 f^{abc}A_\mu^a\left(\bar\varphi_\mu^{bc}+\varphi_\mu^{bc}\right)
+
\frac{m^2}{2}A_\mu^aA_\mu^a-d(N_c^2-1)\gamma^4
\Bigg].
\end{aligned}
\end{equation}

Under the nilpotent BRST variations \cite{Vandersickel:2012tz}
\begin{center}
\begin{tabular}{cc}\label{eq:BRST_def}
    $s A_\mu^a = -D_\mu^{ab}c^b$ & \hspace{2cm}$sc^a=\frac g2f^{abc}c^bc^c$ \\ 
    $s\bar c^a = ib^a$ & $s b^a=0$ \\ 
    $s\varphi_\mu^{ab}=\omega_\mu^{ab}$ & $s\omega_\mu^{ab}=0$\\
$s\bar\omega_\mu^{ab}=\bar\varphi_\mu^{ab}$ & $s\bar\varphi_\mu^{ab}=0$,
\end{tabular}
\end{center}
the RGZ action (\ref{eq:RGZ-action-full}) is not invariant, but displays a soft breaking. A BRST-invariant version of the GZ and RGZ actions has been proposed and many of its consequences are still under investigation \cite{CapriEtAl2015_LCG,CapriEtAl2015_NilpotentBRST,CapriEtAl2016_MoreLCG,CapriEtAl2016_LocalBRST_Ah,CapriEtAl2016_A2min,CapriEtAl2017_RGZ_LCG,Capri:2017bfd,CapriEtAl2017_Nielsen,CapriEtAl2018_GaugeFixings,CapriEtAl2018_Universal,CapriEtAl2018_SYM_Stueckelberg,DudalEtAl2019_BRSTvacuum,CapriSorellaTerin2021_RGZ,DudalVercauteren2023_GapEq}. Such a generalized theory allows one to develop the RGZ theory beyond Landau gauge. Indeed, the Landau gauge action (\ref{eq:RGZ-action-full}) corresponds to the Landau gauge limit of the full BRST invariant RGZ theory we just mentioned. The Landau gauge action (\ref{eq:RGZ-action-full}) is multiplicatively renormalizable 
\cite{Zwanziger1993,Sorella_a_refinement}, and has been 
extensively explored in many contexts, especially in the calculation of correlation functions \cite{Vandersickel:2012tz} and in finite temperature QFT 
\cite{CanforaEtAl2015}. 

As it will become clearer in the next sections,
the quasiparticles of the quadratic RGZ theory 
will be described by field operators which combine
specific components of the gluon field $A_\mu$ and 
of the Zwanziger auxiliary field $\varphi_\mu$.

% ------------------------------------
\section{The emergence of quasiparticles in RGZ}
\label{sec:iparticles}
% ------------------------------------

Now that we have set up the action for the theory, let us 
consider only the quadratic terms of the Landau 
gauge RGZ action, which now reads
\begin{eqnarray}\label{eq:RGZ-quadratic}
    S_{quad} =&& \int\,d^dx
    \left[
      \frac12A_\mu^a \left((-\partial^2+m^2)\delta_{\mu\nu}+\partial_\mu\partial_\nu\right)A_\nu^a + ib^a\partial_\mu A^a_\mu - \bar c^a(-\partial^2)c^a+
    \right.\nonumber\\
    && \left.+\bar\varphi_\mu^{ab}(-\partial^2+M^2)\varphi_\mu^{ab}
    + ig\gamma^2f^{abc}A_\mu^a(\bar\varphi_\mu^{bc}+\varphi_\mu^{bc})
    - \bar\omega_\mu^{ab}(-\partial^2+M^2)\omega_\mu^{ab}
    \right].
\end{eqnarray}

The diagonalization of $S_{quad}$ as a quadratic form
leads to the definition of new fields, which do not possess 
mixed propagators. Such fields (which are linear combinations
of the fields in the Lagrangian) can be used to construct field modes formally similar to 
quasiparticles of the theory. In the context of the original GZ theory (where $m=M=0$), 
such field operators have propagators with imaginary 
poles in the $p^2$ complex plane, corresponding not 
to physical particles, but to the so-called {\it i-particles}.
This feature has been explored in several papers \cite{Sorella_i_particles,Sorella2011IParticles,CapriEtAl2011Glueballs},
especially in connection with the glueball spectrum.

The quadratic action (\ref{eq:RGZ-quadratic}) leads to
the set of (tree-level) propagators \cite{bib:nele,Apollo-thesis}
\begin{eqnarray}\label{eq:RGZ-tree-propagators}
\langle A_\mu^a(p)A_\nu^b(-p)\rangle
&=&
\delta^{ab}\,\frac{p^2+M^2}{(p^2+m^2)(p^2+M^2)+2N_cg^2\gamma^4}\,P^T_{\mu\nu}(p)\equiv\delta^{ab}P_{\mu\nu}^T(p)D_{AA}(p)
\nonumber\\
\langle \bar c^{\,a}(p)c^b(-p)\rangle
&=&
\frac{\delta^{ab}}{p^2}
\nonumber\\
\langle \bar\omega_\mu^{ab}(p)\omega_\nu^{cd}(-p)\rangle
&=&
\frac{\delta^{ac}\delta^{bd}\delta_{\mu\nu}}{p^2+M^2}
\nonumber\\
\langle A_\mu^{a}(p)\varphi_\nu^{bc}(-p)\rangle
=
\langle A_\mu^{a}(p)\bar\varphi_\nu^{bc}(-p)\rangle
&=&
\frac{-i g\gamma^2 f^{abc}}{(p^2+m^2)(p^2+M^2)+2N_cg^2\gamma^4}\,P^T_{\mu\nu}(p),
\nonumber\\
\langle \varphi_\mu^{ab}(p)\varphi_\nu^{cd}(-p)\rangle
=
\langle \bar\varphi_\mu^{ab}(p)\bar\varphi_\nu^{cd}(-p)\rangle
&=&
\frac{-g^2\gamma^4 f^{abm}f^{cdm}}{(p^2+M^2)\,(p^2+m^2)(p^2+M^2)+2N_cg^2\gamma^4}\,P^T_{\mu\nu}(p),
\nonumber\\
\langle \bar\varphi_\mu^{ab}(p)\varphi_\nu^{cd}(-p)\rangle
&=&
\frac{\delta^{ac}\delta^{bd}\delta_{\mu\nu}}{p^2+M^2}
-
\frac{g^2\gamma^4 f^{abm}f^{cdm}}{(p^2+M^2)\,(p^2+m^2)(p^2+M^2)+2N_cg^2\gamma^4}\,P^T_{\mu\nu}(p),
\end{eqnarray}
where $P^T_{\mu\nu}(p)$ is the transverse projector and $D_{AA}(p)$ is the gluon propagator form factor. For explicit loop calculations, it is convenient to write the gluon propagator form factor as a
sum of simple poles, i.e.,
\begin{equation}\label{eq:RGZ-partial-fractions}
  D_{AA}(p)=\frac{R_+}{p^2+a_+^2}+\frac{R_-}{p^2+a_-^2},
\end{equation}
where
\begin{equation}\label{eq:RGZ-residues}
R_{\pm}=\frac12\pm\frac{(m^2-M^2)}{2\Delta},
\qquad
a_{\pm}^2=\frac{(m^2+M^2)\pm\Delta}{2}.
\end{equation}
where we define $\Delta:=\sqrt{(m^2-M^2)^2-8g^2N_c\gamma^4}$.

An obvious feature of this 
set of propagators is the presence of mixed propagators (e.g., $\langle A_\mu^a(p)\varphi_\nu^{bc}(-p)\rangle$),
which comes from the nondiagonal mass matrix in 
(\ref{eq:RGZ-quadratic}). The fields we call "quasiparticles fields" result from the diagonalization of the action (\ref{eq:RGZ-quadratic}) as a quadratic form.

%%%%
\subsection{Splitting the auxiliary fields in adequate color components}\label{subsec:splitting-aux-fields}

In order to find the quasiparticles of the theory, we need 
to diagonalize the quadratic form (\ref{eq:RGZ-quadratic}). We do this by using the same technique of color decomposition in \cite{Sorella_i_particles}, the main difference to this paper being the nonzero mass parameters $m^2$ and $M^2$, typical of RGZ. Having the goal of diagonalization in mind, let us first put the $A_\mu$ and $\varphi_\mu$ fields on an equal footing. First, since we work in Landau gauge, 
we may neglect the terms proportional to 
$\partial_\mu A_\mu^a$, 
which is equivalent to making an appropriate shift in the 
Nakanishi-Lautrup field $b$ (a transformation with trivial 
Jacobian). Next, we note that the mixing between the gluon
field $A$ and the auxiliary fields $\bar\varphi$ and 
$\varphi$ depends only on the sum $\bar\varphi+\varphi$. 
Therefore, we conveniently define the real fields
\begin{eqnarray}\label{eq:U-and-V-def}
V_\mu^{ab}&:=&\frac{\varphi_\mu^{ab}+\bar\varphi^{ab}_\mu}{\sqrt2},\nonumber\\
U_\mu^{ab}&:=&\frac{\varphi_\mu^{ab}-\bar\varphi^{ab}_\mu}{\sqrt2i},
\end{eqnarray}
which can be seen as $V={\sqrt{2}}Re(\varphi)$ and $U={\sqrt{2}}Im(\varphi)$.
With this, the quadratic action (\ref{eq:RGZ-quadratic})
is written as
\begin{eqnarray}\label{eq:RGZ-quad-U-V}
    S_{quad} \rightarrow&& \int\,d^dx
    \left[
      \frac12A_\mu^a \left(-\partial^2+m^2\right)A_\mu^a + ib^a\partial_\mu A^a_\mu - \bar c^a(-\partial^2)c^a
      - \bar\omega_\mu^{ab}(-\partial^2+M^2)\omega_\mu^{ab}+
    \right.\nonumber\\
    && \left.+\frac12 U_\mu^{ab}(-\partial^2+M^2)U_\mu^{ab}
    +\frac12 V_\mu^{ab}(-\partial^2+M^2)V_\mu^{ab}
    + ig\sqrt{2}\gamma^2f^{abc}A_\mu^a V_\mu^{bc}
    \right].
\end{eqnarray}

Note that $U$, the imaginary part of $\varphi$, decouples 
from other fields,
while $V$ does not. Furthermore, the quadratic coupling 
between $A^a$ and $V^{bc}$ contains the $SU(N_c)$ structure 
constants $f^{abc}$. This means that the color-symmetric 
part of the field $V^{bc}$ does not couple directly to $A$,
while their color-antisymmetric components do. 

With this in mind, one might be tempted to simply assume that
the color tensor $V^{ab}$ is antisymmetric, since its 
color-symmetric components would be canceled in the 
path-integration of the fermionic auxiliary fields. One can straightforwardly check that this is not true for
the general case of interacting fields (although this
assumption works if one were restricted to the quadratic action). 
A more generally valid approach is to decompose the field
$V$ into color-symmetric and color-antisymmetric components,
as in \cite{Sorella_i_particles}, so that
\begin{eqnarray}\label{eq:decomposition_V}
 V_\mu^{ab} =: \sigma_\mu^{ab} +  \frac{f^{abc}V_\mu^c}{\sqrt{N_c}} + \alpha_\mu^{ab},   
\end{eqnarray}
where we define the color-symmetric field
\begin{eqnarray}\label{eq:symm-sigma}
    \sigma_\mu^{ab}&:=&\frac{V_\mu^{ab}+V^{ba}_\mu}{2},
\end{eqnarray}
while the color-antisymmetric component is conveniently written 
as 
\begin{eqnarray}\label{eq:decomp-V-antisym}
    \frac{V_\mu^{ab}-V^{ba}_\mu}{2}=:\frac{f^{abc}V_\mu^c}{\sqrt{N_c}} + \alpha_\mu^{ab}.
\end{eqnarray}
By construction, $\alpha_\mu$ is a 
color-antisymmetric tensor such that
$f^{abc}\alpha_\mu^{bc}=0$, and $V_\mu^c$ is a 
vector field with one color index only. 
Note that the decomposition
(\ref{eq:decomp-V-antisym}) can be interpreted as a sum
of a component ``parallel'' to $f^{abc}$ and another 
component ``orthogonal'' to it. Note that, if needed, one may write explicitly
\begin{eqnarray}
        \alpha_\mu^{ab} := \left(\delta^{am}\delta^{bn} - \frac1{N_c}f^{abc}f^{cmn}\right)\left(\frac{V_\mu^{mn}-V^{nm}_\mu}{2}\right).
\end{eqnarray}

In terms of these variables, the quadratic action 
(\ref{eq:RGZ-quad-U-V}) now reads
\begin{eqnarray}\label{eq:RGZ-action-quasiparticles}
    S_{quad} \rightarrow&& \int\,d^dx
    \left[
      \frac12A_\mu^a \left(-\partial^2+m^2\right)A_\mu^a + ib^a\partial_\mu A^a_\mu - \bar c^a(-\partial^2)c^a
      - \bar\omega_\mu^{ab}(-\partial^2+M^2)\omega_\mu^{ab}
      +\frac12 U_\mu^{ab}(-\partial^2+M^2)U_\mu^{ab}+
    \right.\nonumber\\
    && \left.
    +\frac12 \sigma_\mu^{ab}(-\partial^2+M^2)\sigma_\mu^{ab}
    +\frac12 \alpha_\mu^{ab}(-\partial^2+M^2)\alpha_\mu^{ab}
    +\frac12 V_\mu^{a}(-\partial^2+M^2)V_\mu^{a}
    + ig\sqrt{2N_c}\gamma^2A_\mu^a V_\mu^{a}
    \right]\nonumber\\
    &=& \int\,d^dx
    \left[
      \frac12A_\mu^a \left(-\partial^2+m^2\right)A_\mu^a + ib^a\partial_\mu A^a_\mu
    +\frac12 V_\mu^{a}(-\partial^2+M^2)V_\mu^{a}
    + ig\sqrt{2N_c}\gamma^2A_\mu^a V_\mu^{a}  
    \right] +    S_{quad}^{(others)},
\end{eqnarray}
where we used the (anti-)symmetry properties of $\sigma$ and
$\alpha$, as well as $f^{abc}f^{abd}=N_c\delta^{cd}$ and
\begin{eqnarray}
    S_{quad}^{(others)}&=&
    \int d^dx \left[
    \frac12 U_\mu^{ab}(-\partial^2+M^2)U_\mu^{ab}
    +\frac12 \sigma_\mu^{ab}(-\partial^2+M^2)\sigma_\mu^{ab}
    +\frac12 \alpha_\mu^{ab}(-\partial^2+M^2)\alpha_\mu^{ab}\right.\nonumber\\
    &&\left.- \bar c^a(-\partial^2)c^a
    - \bar\omega_\mu^{ab}(-\partial^2+M^2)\omega_\mu^{ab}
    \right].
\end{eqnarray}
In this form, the only nondiagonal coupling left 
(besides the Landau gauge constraint) is
the one between $A^a_\mu$ and $V^a_\mu$ (i.e., the gluon and the component of $\varphi$ which is 
real, antisymmetric and proportional to the structure 
constant). 
This mixing can be lifted by a rotation by an imaginary angle 
$i\theta$ in the field subspace $(A,V)$, which can be 
implemented by defining the new fields
\begin{eqnarray}\label{eq:quasi-particles}
    \lambda_\mu^a&:=&A_\mu^a\cosh\theta + iV_\mu^a\sinh\theta\nonumber\\
    \eta_\mu^a&:=&-iA_\mu^a\sinh\theta + V_\mu^a\cosh\theta,
\end{eqnarray}
where
\begin{eqnarray}\label{eq:theta}
    \tanh(2\theta)=\frac{2\sqrt{2N_c}g\gamma^2}{m^2-M^2}.
\end{eqnarray}
Since the linear transformation (\ref{eq:quasi-particles}) 
has Jacobian $\cosh^2\theta-\sinh^2\theta=1$, the 
functional measure is unchanged.

With these transformations, the quadratic action 
(\ref{eq:RGZ-quadratic}) reads
\begin{eqnarray}\label{eq:quad-action-nontransverse}
    S_{quad}=\int\,d^dx\,\left[
        \frac12\lambda_\mu^a(-\partial^2+M_\lambda^2)\lambda_\mu^a + ib^a\partial_\mu(\lambda_\mu^a\cosh\theta - i\eta_\mu^a\sinh\theta) +
        \frac12\eta_\mu^a(-\partial^2+M_\eta^2)\eta_\mu^a
    \right] + S_{quad}^{(others)},
\end{eqnarray}
where the massive parameters can be written as 
\begin{eqnarray}\label{eq:M-lambda_M-eta}
    M_\lambda^2&:=& \frac{m^2\cosh^2\theta+M^2\sinh^2\theta}{\cosh(2\theta)}
    =\frac{m^2+M^2+{\rm sgn}(m^2-M^2)\sqrt{(m^2-M^2)^2-8N_cg^2\gamma^4}}{2},
    \nonumber\\
    M_\eta^2&:=& \frac{M^2\cosh^2\theta+m^2\sinh^2\theta}{\cosh(2\theta)}=\frac{m^2+M^2-{\rm sgn}(m^2-M^2)\sqrt{(m^2-M^2)^2-8N_cg^2\gamma^4}}{2},
\end{eqnarray}
where ${\rm sgn}(x):=x/|x|$.

Let us remark that mass poles in the negative real axis (so that $M_\lambda^2\in\mathbb{R}_+$ and $M_\eta^2\in\mathbb{R}_+$) are supported by the model and their appearance corresponds to three conditions. Namely, one must have $\theta\in\mathbb{R}$ (or, equivalently, $(m^2-M^2)^2-8g^2N_c\gamma^4>0$), and $M_\lambda^2 M_\eta^2=m^2M^2+2N_cg^2\gamma^4> 0$, and also $m^2+M^2>0$. In sum, if real poles are present, then the inequality $-m^2M^2 < 2g^2N_c\gamma^4 < (m^2-M^2)^2/4$ must be respected. In the case $m^2<0$ and $M^2>0$ (which is favored in lattice best fits for the Landau gauge gluon propagator \cite{DudalOliveiraSilva2018}), the window in parameter space for real poles can become quite narrow. Indeed, the best fit for a $64^4$ lattice in \cite{DudalOliveiraSilva2018} is such that $M^2_{Latt}=2.521\, {\rm GeV}^2$, $m^2_{Latt}=-2.013\, {\rm GeV}^2$, and $2g^2N_c\gamma^4_{Latt}=5.354\, {\rm GeV}^4$. With these parameters, the reality condition is not satisfied, but a relatively small change in the parameters may lead to real mass poles. More specifically, one has $(m^2_{Latt}-M^2_{Latt})^2= 20.56\, {\rm GeV}^4$ and $8g^2N_c\gamma^4_{Latt}=21.42{\rm GeV}^4$, a difference of only $4.2\%$. Thus, in some sense, one may think that the lattice best fit (with complex masses) is not very far away from the real mass case. In this paper, while we avoid difficulties involving complex masses (such as a proper definition of a physical Hilbert space) by assuming real mass poles, we also note that propagators with real poles provide reasonable fits for the Landau gauge gluon propagator.

Note that (\ref{eq:quad-action-nontransverse}) still 
has a mixing
term involving both quasiparticle field operators, which is 
just a rephrasing of the Landau gauge condition $\partial_\mu A_\mu=0$, written in terms of 
$\lambda_\mu^a$ and $\eta_\mu^a$. By calculating the propagators associated with this quadratic
action (\ref{eq:quad-action-nontransverse}), one finds 
a nonzero mixed propagator 
$\langle\lambda_\mu^a \eta_\mu^a\rangle$. Besides, the 
$\langle{\lambda_\mu^a\lambda_\nu^b}\rangle$
and $\langle{\eta_\mu^a\eta_\nu^b}\rangle$
propagators have longitudinal parts. This signals that 
the RGZ action needs further information in order to be 
properly diagonalized.

%%%%%%%%
\subsection{Transverse auxiliary fields and 
the diagonalization of the quadratic action}

Before going further, let us now discuss how the 
transversality of the Zwanziger auxiliary fields is 
connected to 
a meaningful definition of quasiparticles.

%%%%%
\subsubsection{Original RGZ}

We start considering how the original GZ and RGZ 
theories may lead to quasiparticle field 
operators without mixed propagators. As it turns out, 
the quadratic form $S_{quad}$ may be diagonalized,
but the resulting field operators will be defined 
using the transverse projector 
$P^T_{\mu\nu}=\delta_{\mu\nu}-\partial_\mu\partial_\nu/\partial^2$
and the longitudinal projector $P^L_{\mu\nu}=\partial_\mu\partial_\nu/\partial^2$, 
which are nonlocal operators. Indeed, let us define
$\lambda^T:=\cosh\theta\,(P^TA)+i\sinh\theta (P^TV)$, 
$\eta^T:=-i\sinh\theta\,(P^TA)+\cosh\theta (P^TV)$, and 
$V_L=(P^LV)$. With these transformations, the $\langle\lambda^T\lambda^T\rangle$
and $\langle\eta^T\eta^T\rangle$ propagators are both transverse, and $\langle\lambda^T\eta^T\rangle=0$. Therefore, the quadratic action is diagonalizable (as any quadratic form), but only in terms of the nonlocal field operators $\lambda^T$ and $\eta^T$.
Another possibility in the original RGZ theory is to 
simply define the would-be quasiparticles as in 
(\ref{eq:quasi-particles}). As a consequence, the tree-level
propagators $\langle\lambda\lambda\rangle$
and $\langle\eta\eta\rangle$ are not transverse and, 
besides, $\langle\lambda\eta\rangle\not=0$. 

%%%%%
\subsubsection{Transverse RGZ}

Since the original RGZ theory leads not only to nonzero 
longitudinal parts of the quasiparticle two-point 
functions, but also to an unwanted mixed propagator
between them, we now explore another possibility. The very transversality of the nonlocal operators $\lambda^T$ 
and $\eta^T$ suggests that it may be interesting  
to impose that the quasiparticle fields are themselves 
transverse as a constraint of the theory. In order 
to do so, let us assume that the auxiliary fields 
$(\bar\varphi,\varphi,\bar\omega,\omega)$ are transverse 
themselves.
It is possible to impose this condition through the 
introduction of two BRST doublets, namely 
$(\bar c_\omega,\bar b_\varphi)$ and $(\bar b_{\bar\varphi},c_{\bar\omega})$, so that
\begin{eqnarray}
    s\bar c_\omega^{ab} &=& i\bar{b}_\varphi^{ab}\nonumber\\
    s\bar{b}_\varphi^{ab}&=&0
\end{eqnarray}
and
\begin{eqnarray}
    sb_{\bar\varphi}^{ab} &=& ic_{\bar\omega}^{ab}\nonumber\\
    sc_{\bar\omega}^{ab}&=&0.
\end{eqnarray}

The transversality of the auxiliary fields is then introduced by adding the BRST exact term 
\begin{eqnarray}\label{eq:transversality-varphi}
S_{\text{Tr}\varphi} &=& \int d^dx \, s\left(
    \bar c_{\omega}^{ab} \partial_\mu \varphi^{ab}_\mu
    +  ib_{\bar\varphi}^{ab} \partial_\mu \bar\omega^{ab}_\mu
    \right) \\\nonumber
    &=& \int d^dx \, \Bigl[ i\bar{b}_\varphi^{ab} (\partial_\mu \varphi_\mu^{ab})
    + ib_{\bar\varphi}^{ab} (\partial_\mu \bar\varphi_\mu^{ab})
    - \bar c_\omega^{ab} (\partial_\mu \omega_\mu^{ab})
    - c_{\bar\omega}^{ab} (\partial_\mu \bar\omega_\mu^{ab}) \Bigr],
\end{eqnarray}
to the action, where we used the usual BRST transformations of the 
auxiliary fields, namely
\begin{eqnarray}
    s\varphi_\mu^{ab}&=&\omega_\mu^{ab}\nonumber\\
    s\omega_\mu^{ab}&=&0\nonumber\\
    s\bar\omega_\mu^{ab}&=&\bar\varphi_\mu^{ab}\nonumber\\
    s\bar\varphi_\mu^{ab}&=&0,
\end{eqnarray}
so that the RGZ action becomes $S_{RGZ}\rightarrow S_{RGZ}+S_{Tr\varphi}$.
Note that (\ref{eq:transversality-varphi}) is such that
all auxiliary fields 
$(\varphi,\bar\varphi,\omega,\bar\omega)$ are constrained
to be transverse. It can be easily shown that this condition
does not alter the quadratic part of the horizon function which is obtained 
after the functional integration of all auxiliary fields. However, the interaction terms are changed when transversality is imposed on the auxiliary fields. A detailed comparison between the original and the transverse RGZ models beyond the quadratic approximation should include the Ward identities of the model, its renormalizability and the possible changes to the Gribov-Zwanziger horizon function in the interacting case. Since we are interested now only in the quadratic approximation of the theory, this analysis is beyond the scope of this paper and will be made elsewhere.

If one rewrites the transversality constraint on
$(\bar\varphi,\varphi)$ in terms of the real and imaginary part fields $(V,U)$, then
\begin{eqnarray}
    S_{Tr\,UV}=\int\,d^dx\,\left[
     ib_V^{ab}\partial_\mu V_\mu^{ab}-ib_U^{ab}\partial_\mu U_\mu^{ab}     - \bar c_\omega^{ab} (\partial_\mu \omega_\mu^{ab})
    - c_{\bar\omega}^{ab} (\partial_\mu \bar\omega_\mu^{ab})
    \right]
\end{eqnarray}
is simply the transversality constraint of $\bar\varphi$ 
and $\varphi$ written in terms of $U$ and $V$, with
\begin{eqnarray}
    b_V^{ab}&:=&\frac{\bar b_\varphi^{ab}+b_{\bar\varphi}^{ab}}{\sqrt2}\nonumber\\
    b_U^{ab}&:=&\frac{\bar b_\varphi^{ab}-b_{\bar\varphi}^{ab}}{i\sqrt2}.
\end{eqnarray}

If one proceeds with the same decompositions in color
space as 
in Subsection \ref{subsec:splitting-aux-fields}, 
one eventually finds the part of the quadratic action  relevant for the quasiparticles
\begin{eqnarray}\label{eq:transverse-quad-action}
    S^{\lambda\eta}_{quad}=\int\,d^dx\,\left[
        \frac12\lambda_\mu^a(-\partial^2+M_\lambda^2)\lambda_\mu^a + ib_\lambda^a\partial_\mu\lambda_\mu^a +
        \frac12\eta_\mu^a(-\partial^2+M_\eta^2)\eta_\mu^a + ib_\eta^a\partial_\mu\eta_\mu^a 
    \right],
\end{eqnarray}
with the corresponding Nakanishi-Lautrup fields
\begin{eqnarray}\label{eq:quasi-part-transv}
    b_\lambda^a&:=&b^a\cosh\theta + ib_V^a\sinh\theta\nonumber\\
    b_\eta^a&:=&-ib^a\sinh\theta + b_V^a\cosh\theta.
\end{eqnarray}

Now it is clear that the quadratic action 
(\ref{eq:transverse-quad-action}) 
provides the transverse propagators for the quasiparticles
\begin{eqnarray}\label{eq:propagators-quasiparticles}
    \langle{\lambda_\mu^a(p)\lambda_\nu^b(-p)}\rangle&=&
    \frac{\delta^{ab}}{p^2+M_\lambda^2}\left(\delta_{\mu\nu}-\frac{p_\mu p_\nu}{p^2}\right)\nonumber\\
    \langle{\eta_\mu^a(p)\eta_\nu^b(-p)}\rangle&=&\frac{\delta^{ab}}{p^2+M_\eta^2}\left(\delta_{\mu\nu}-\frac{p_\mu p_\nu}{p^2}\right)
\end{eqnarray}
and, as advertised, the tree-level mixed propagator $\langle \lambda_\mu^a(p)\eta_\nu^b(-p)\rangle=0$. The spectral functions corresponding to these propagators are simply
\begin{eqnarray}\label{eq:quasiparticle-spec-func}
    \rho_{\lambda\lambda}(s)&=&\delta(s-M_\lambda^2)\nonumber\\
    \rho_{\eta\eta}(s)&=&\delta(s-M_\eta^2),
\end{eqnarray}
which are nonnegative for real positive $M_\lambda^2,M_\eta^2\in\mathbb{R}$. For comparison, the gluon spectral function reads 
\begin{eqnarray}\label{eq:gluon-spec-func}
    \rho_{AA}(s)=\cosh^2\theta\,\delta(s-M_\lambda^2)-\sinh^2\theta\,\delta(s-M_\eta^2),
\end{eqnarray}
which is not positive-definite, but still obeys the sum rule $\int_0^\infty ds\,\rho_{AA}(s)=1$.

In a generic situation, the mass poles $M_\lambda$ and $M_\eta$ can be complex or real, depending precisely on the reality (or not) of the angle $\theta$ in (\ref{eq:theta}). In the case of real poles (so that $\theta\in\mathbb{R}$), the residues (\ref{eq:RGZ-residues}) of the original fields $A_\mu$ and $V_\mu$ are also real, but one of them is negative, and the other is larger than one. On the other hand, the residues of the quasiparticle propagators (\ref{eq:propagators-quasiparticles}) are equal to $1$ and simply correspond to free, massive particles in the quadratic approximation.
In the present paper, we shall mainly consider this 
case, leaving the details of the complex mass case
for future studies.
These results are, as could be anticipated, in line 
with the quantum-mechanical and scalar QFT toy models 
in \cite{Mintz:2024soo}, whose analysis can be seen as a preparation to the one in this paper. We consider the present work as a first step of generalization of that previous work to a non-abelian gauge theory.

% ------------------------------------
\section{Pseudohermiticity as a tool to interpret the RGZ quasiparticles}\label{sec:pseudohermiticity}
% ------------------------------------

Let us now take a step further towards the interpretation of the quasiparticles in the RGZ theory. For this, let us invoke a symmetry of the action, which has been gaining relevance in the past years, the so-called ${\cal PT}$-symmetry \cite{Bender_1998}. After a quick review of its basics, we notice that the RGZ action is ${\cal PT}-$symmetric and then we discuss a few consequences of this fact.

%%%%%%
\subsection{${\cal PT}$-symmetry and pseudohermiticity: a quick review}\label{sec:PTsymm-review}

According to the von Neumann axioms of Quantum Mechanics, observables correspond to self-adjoint operators \cite{vonNeumann1927MathematicalFoundation,vonNeumannQMbook}. Crucial reasons for that are the reality of the spectrum of such operators and the orthogonality of eigenvectors associated with different eigenvalues, plus their completeness, expressed in terms of the spectral theorem. Furthermore, the fact that the Hamiltonian operator is self-adjoint implies a unitary time evolution operator, which is a direct application of Stone's theorem \cite{Stone1932}.

In 1998, however, a paper by Carl Bender and Stefan Boettcher opened the possibility for a somewhat weaker constraint \cite{Bender_1998}. In that work, the authors showed numerically that a wide class of nonhermitian quantum mechanical hamiltonians had a completely real point spectrum, a result which generalized previous result for specific potentials \cite{BenderTurbiner1993}. In the following years, rigorous results confirmed their findings in specific (but still interesting) cases, such as the one-dimensional potentials $V(x)=igx^3$ \cite{Shin_2002_229} and $V(x)=-gx^4$ (with $g>0$) \cite{Dorey_2001_34} (both potentials have real spectra, bounded from below).

Of course, the orthogonality of the set of eigenvectors of a nonhermitian operator is generally not satisfied, even when eigenvalues are real. In order to recover the orthonormality of the set of eigenvectors, the inner product of the Hilbert space has to be redefined \cite{Bender_2002_89}. It should be then possible to define a new Hilbert space, where the Hamiltonian becomes a self-adjoint operator and time evolution, unitary \cite{Mannheim:2009zj,Mostafazadeh_2010_07}.

The reason behind the reality of the spectrum has been identified with a symmetry of the action, which was called a ${\cal PT}-$symmetry, due to its similarities (but not literal identification) with the spacetime parity and time-reversal symmetries \cite{Sakurai,Bender_1998}. More specifically, there is an antilinear involution operator ${\cal PT}=({\cal PT})^\dagger$ with $({\cal PT})^2=1$ such that $[\hat H,{\cal PT}]=0$, where $\hat H$ is the Hamiltonian operator of a given theory.\footnote{Note that, in many cases (including the one of this paper), the operators ${\cal P}$ and ${\cal T}$ have little to do with actual parity and time inversion, except for the algebraic properties just listed.} A few years later, Ali Mostafazadeh showed that whenever a ${\cal PT}-$symmetric Hamiltonian $\hat H$ has real spectrum, there is a positive, invertible and bounded operator $\hat\eta$ which makes it $\eta-$pseudohermitian, i.e., $\hat H^\dagger=\hat\eta \hat H\hat\eta^{-1}$ (at least in a finite-dimensional Hilbert space) \cite{Mostafazadeh_2002_435}. For this reason, we call such Hamiltonians ${\cal PT}-$symmetric or pseudohermitian interchangeably. Very importantly, it also turns out that, this very operator $\hat\eta$ is the one to be used as a metric operator, in order to define the new Hilbert space of the theory, which renders the Hamiltonian self-adjoint. (Of course, this is only possible if its spectrum is real.) Let us now explain this crucial point in more detail.

The Hilbert space ${\cal H}$ where the theory is initially defined is called the {\it reference} Hilbert space. Its inner product is denoted by $\langle\cdot|\cdot\rangle$ in the usual Dirac notation. On the other hand, one may define a different inner product (thus a different Hilbert space ${\cal H}_\eta$) given by $\langle\cdot|\hat\eta\cdot\rangle\equiv\langle\cdot|\cdot\rangle_\eta$. We denote $\hat A^\dagger$ the adjoint of $\hat A$ in the Hilbert space ${\cal H}$, whereas $\hat A^\sharp$ is the adjoint of $\hat A$ in ${\cal H}_\eta$. More explicitly, we define $\hat A^\sharp$ by the equality $\langle\psi|\hat A\varphi\rangle_\eta
    =: \langle\hat A^\sharp\psi|\varphi\rangle_\eta$. Given the defining relation $\langle\cdot|\hat\eta\cdot\rangle\equiv\langle\cdot|\cdot\rangle_\eta$, which connects both inner products, one immediately has $\hat A^\sharp=\hat\eta^{-1}\hat A^\dagger\hat\eta$. 
    An immediate and important consequence of this relation is that any $\eta-$pseudohermitian $\hat A=\hat\eta^{-1}\hat A^\dagger\hat\eta$ is actually hermitian in the $\eta$-inner product, i.e., $\hat A^\sharp=\hat A$, even though it is not hermitian in the inner product of the reference Hilbert space ${\cal H}$. We say that such operators with $\hat A^\sharp=\hat A$ are $\eta-$pseudohermitian \cite{Mostafazadeh_2010_07,Mostafazadeh_2010_82}.

In general, the spectrum of a generic pseudohermitian operator is composed of real values, plus pairs of complex-conjugate values \cite{Mostafazadeh_2002_431}. This structure will be crucial for our analysis. Eigenvectors of the Hamiltonian which have real eigenvalues also correspond to eigenvectors of the antilinear ${\cal PT}$ operator. However, eigenvectors associated with complex eigenvalues are not eigenstates of ${\cal PT}$. This is called a ${\cal PT}$-symmetry breaking, given its similarity with spontaneous symmetry breaking (i.e., such complex energy states do not share the ${\cal PT}$ symmetry of the Hamiltonian). Therefore, the set of complex energy eigenstates is often called a ${\cal PT}$-broken phase \cite{Mostafazadeh_2002_431,Mostafazadeh_2002_435,Mostafazadeh_2002_438}.

In the ${\cal PT}$-broken phase, the spectrum has complex elements and the metric operator $\hat\eta$ is no longer positive \cite{Mostafazadeh_2010_82}. This means that it may not be used as a metric operator and the Hamiltonian will not be Hermitian. This is an important complication which has to be dealt with in the ${\cal PT}$-broken phase in order to find a proper definition of a physical Hilbert space. Even though we acknowledge that such regime of complex energies is the most interesting one for the discussion of the gluon propagator (given the complex mass poles which appear in fits of lattice Yang-Mills calculations), we will leave an explicit discussion for  future investigation. In this paper, we will only consider explicitly the case of real energies (thus positive metric operator). Note also that real mass poles do not yield the best fit for the gluon propagator, but the corresponding results are not very far away from lattice points \cite{PelaezPalharesBarriosMintz-2026}.

In the case of real spectrum, the metric operator $\hat\eta$ is a positive operator. It then follows that it has a positive and hermitian square root $\hat\rho=\hat\eta^{1/2}=\hat\rho^\dagger$. This operator can be used to define observables \cite{Mostafazadeh_2010_07}. Given an operator $\hat o=\hat o^\dagger\not=\hat o^\sharp$ (Hermitian in ${\cal H}$, but not in ${\cal H}_\eta$), one defines $\hat O:=\hat\rho^{-1}\hat o\hat\rho$ which is clearly $\eta-$pseudohermitian, i.e., $\hat O^\sharp=\hat O$. Therefore, the operator $\hat O$ is (at a least a candidate for) an observable operator in ${\cal H}_\eta$. Another very important role of the operator $\hat\rho$ is to obtain a hermitian operator $\hat h$ from the original Hamiltonian $\hat H$, whenever ${\rm{Spec}}(\hat H)\subseteq\mathbb{R}$. Indeed, one may define $\hat h:=\hat \rho\hat H\hat\rho^{-1}=\hat h^\dagger$ whenever $\hat H$ is $\eta-$pseudohermitian. The operator $\hat h$  is called a hermitian equivalent Hamiltonian. 

The fact that the Hamiltonian is not self-adjoint in the reference Hilbert space also implies interesting consequences for observables. A particularly interesting feature is more clearly seen in the Heisenberg picture. As discussed in \cite{Mintz:2024soo}, an operator $\hat x$ which is hermitian at time $t_0$ and evolves under a nonhermitian Hamiltonian according to the Heisenberg equation, will not remain hermitian. In other words, $\hat x(t)\not=[\hat x(t)]^\dagger$ for $t\not=t_0$, even if $\hat x(t_0)=[\hat x(t_0)]^\dagger$, in general. Therefore, a consistent definition of an observable operator must be such that it is Hermitian with respect to the $\eta$-inner product. Indeed, it follows from the Heisenberg equation that, if $\hat A=\hat A^\sharp$ at $t=t_0$ and $\hat H=\hat H^\sharp$, then $\hat A(t)=\hat A(t)^\sharp$ for all $t$ \cite{Mintz:2024soo}. 

This observation has direct implications for the study of correlation functions. In \cite{Mintz:2024soo}, the authors recalled explicitly how two-point correlation functions of nonhermitian operators are in general complex and thus violate the Osterwalder-Schrader positivity condition. On the other hand, if hermiticity is recovered by a proper redefinition of the Hilbert space (${\cal H}\rightarrow{\cal H}_\eta$) and also of the observable operators (as discussed above), positivity of the spectral function is recovered. This can be seen from the following argument.

As discussed in \cite{Mintz:2024soo},
Heisenberg-picture operators that are Hermitian in the reference Hilbert space ${\cal H}$ at some initial time $t_0$ become in general nonhermitian for $t\not=t_0$, due to $\hat H^\dagger\not=\hat H$. As a consequence, two-point correlation functions involving such operators will have, in general, a 
complex spectral function. Indeed, it can be shown that, for a Schroedinger picture operator $\hat x=\hat x^\dagger$, the spectral function\footnote{Note that we consider 
the correlation functions as defined in ${\cal H}_\eta$, so that we use the inner product $\langle\cdot|\cdot\rangle_\eta$, since we see this as crucial to recover unitarity.} can be written in the form
\begin{eqnarray}\label{eq:spectral-function}
    \rho^{xx}(s) = \sum_m\frac{1}{2\pi}
    \braopket{\psi_0}{\hat x(0)}{m}_\eta \braopket{\psi_0}{\hat x^\#(0)}{\,m}_\eta^*
    \delta\left[ s - \left(\frac{E_m-E_0}{\hbar}\right)^2\right],
\end{eqnarray}
where $\ket{m}$ are the energy eigenstates of the theory and $E_m$ their respective energies. This means that if $\hat x^\sharp=\hat x$, then $\rho^{xx}(s)\geq0$. Equivalently, if $\rho^{xx}(s)<0$ for some $s$, then $\hat x^\sharp\not=\hat x$. On the other hand, by using the construction above, one may define observable operators of the type $\hat X=\hat \rho^{-1}\hat x\hat\rho$, which are hermitian in ${\cal H}_\eta$ (i.e., $\hat X^\sharp=\hat X$) and thus have a positive spectral function.

In the next section, we shall explore how these concepts may be employed, in a form generalized to QFT, as a tool to interpret the correlation functions involving the quasiparticle fields of RGZ (at least in the regime of real mass poles).

%%%%%%%
\subsection{Possible signs of pseudohermiticity in the strong interactions}\label{sec:pseudohermiticity_strong_int}

Let us start this subsection with a disclaimer. Most of the developments of pseudohermitian quantum systems have been made in the context of Quantum Mechanics, and not much in Quantum Field Theories, at least at the level of strict mathematical proofs. (Of course, many works with a more practical spirit have been put forward in the literature. A short list of such works, which are many, being
\cite{
Alexandre:2017foi,
Alexandre:2018xyy,
Alexandre_2020,
Beygi_2019,
BenderHassanpourKlevanskySarkar2018,
Sablevice_2024_109,
mason2023flavour,
Romatschke:2022jqg,
Lawrence:2023woz,
Mintz:2024soo,
kuntz2024unitarityptsymmetryquantum}.) However, we believe that many ideas in Quantum Mechanics will likely find an analogy in Quantum Field Theories, and their proofs could be eventually found. With this conjectural spirit, we seek to extend some concepts of Pseudohermitian Quantum Mechanics not only to Quantum Field Theories, but actually to a gauge theory.

As discussed in the previous subsection, the spectrum of pseudohermitian Hamiltonians is composed of real energies and, possibly, pairs of complex-conjugate energies. Back to gauge theory, recall that it is a well-established claim that the gluon propagator has no positive-definite K\"allen-Lehmann spectral function, a fact which is often related to confinement \cite{Alkofer:2000wg,Cucchieri_2005,Alkofer_2004,AguilarBinosiPapavassiliou2008,FischerMaasPawlowski2009,Cyrol:2018_SciPost,Dudal_2014,DudalOliveiraSilva2018,2013MPLA...2830035C}. While some authors support that this is due to the presence of pairs of complex-conjugate poles in the momentum $(p^2)$ complex plane \cite{
CucchieriDudalMendesVandersickel2012,
FalcaoOliveiraSilva2020,
Kondo_2020,
deBritoPereira2024,
Fischer:2020xnb}, others explore the possibility that no mass pole exists at all (i.e., the gluon propagator is an entire function) \cite{RobertsWilliamsKrein1992,RobertsWilliams1994,ElBennichKreinRojasSerna2016,Mandula1999GluonPropagator}. In the matter sector, it turns out that the quark propagator also violates reflection positivity and has a complicated analytic structure \cite{BurdenRobertsWilliams1992,Alkofer_2004,AguilarCardonaFerreiraPapavassiliou2018,AguilarFerreiraOliveiraPapavassiliouTeixeira2024,OliveiraFredericoDePaula2025,
HorakPawlowskiWink2023,
FalcaoOliveira2022,
PawlowskiWessely2025,
AlkoferEtAl2026}. In any case, it is still somewhat mysterious how the real classical Yang-Mills action, once quantized, may give rise to such strange properties. Regardless of the mechanism giving rise to this behavior, we believe that the positivity violation of the gluon spectral function in the nonperturbative regime of QCD may be connected to a symmetry of the effective action akin to ${\cal PT}-$symmetry. The present discussion offers a few first thoughts towards the investigation of this hypothesis. 

In this scenario, positivity violation (with or without complex mass poles) would arise from a nonhermitian effective action. In the usual formulation of classical $SU(N_c)$ Yang-Mills action, one defines the gluon field $A_\mu$ as real, so that the Schroedinger picture field $\hat A^\dagger_\mu=\hat A_\mu$.  However, as many nonperturbative methods consistently
show, the spectral function associated to the gluon propagator $\langle A_\mu^a(x)A_\nu^b(y)\rangle$ displays positivity violation. The underlying reason for this would be the hermiticity breaking of the gauge field Heisenberg-picture operator $\hat A_\mu^a(x)$ under time evolution. Such hermiticity breaking would be the result of a pseudohermitian effective action, in analogy to the discussion in \ref{sec:PTsymm-review}.

At this point, the arguments given above may seem too abstract and far from concrete examples in the context of Yang-Mills theories. In order to make the points above more explicit,  let us consider the RGZ action (\ref{eq:RGZ-action-full}) as an effective model for the strong interactions. First, we note that both the RGZ action $S_{RGZ}$ of (\ref{eq:RGZ-action-full}) and its transverse version $S_{RGZ}+S_{Tr\varphi}$ are not just complex but rather ${\cal PT}-$symmetric\footnote{In the sense of an antilinear involution symmetry transformation discussed in the previous subsection, i.e., $({\cal PT})^\dagger={\cal PT}$ and $({\cal PT})^2=1$. Note that it is not an actual parity-time transformation}. Let us define the antilinear field and spacetime transformations\footnote{In this subsection, we omit the caret $( \,\hat{}\, )$ symbol on top of operators, in order to simplify notation.}
\begin{eqnarray}
    \begin{array}{cc}
        {\cal PT}\partial_\mu{\cal PT}=-\partial_\mu & {\cal PT}i{\cal PT}=-i \\
        {\cal PT}A_\mu^a(x){\cal PT}=-A_\mu^a(-x) &
        {\cal PT}b^a(x){\cal PT}=-b^a(-x) \\
        {\cal PT}\bar c^a(x){\cal PT}=\bar c^a(-x) &
        {\cal PT} c^a(x){\cal PT}=c^a(-x) \\
        {\cal PT}\varphi_\mu^{ab}(x){\cal PT}=\varphi_\mu^{ab}(-x) & 
        {\cal PT}\bar\varphi_\mu^{ab}(x){\cal PT}=\bar\varphi_\mu^{ab}(-x)
        \\
        {\cal PT}\omega_\mu^{ab}(x){\cal PT}=\omega_\mu^{ab}(-x) &
        {\cal PT}\bar\omega_\mu^{ab}(x){\cal PT}=\bar\omega_\mu^{ab}(-x)\\
        {\cal PT}b_{\bar\varphi}^{ab}(x){\cal PT}=b_{\bar\varphi}^{ab}(-x) &
        {\cal PT}{\bar b}_{\varphi}^{ab}(x){\cal PT}={\bar b}_{\varphi}^{ab}(-x)
        \\
        {\cal PT}{\bar c}_\omega^{ab}(x){\cal PT}=-{\bar c}_\omega^{ab}(-x) &
        {\cal PT}c_{\bar\omega}^{ab}(x){\cal PT}=-c_{\bar\omega}^{ab}(-x). %
    \end{array}    
\end{eqnarray}
Under these transformations,
 $S_{RGZ}$ is $\mathcal{PT}-$symmetric term by term, i.e.,
\begin{eqnarray}
    {\cal PT}S_{RGZ}{\cal PT}=S_{RGZ}.
\end{eqnarray}

The functional measure is also invariant and also the ${\cal PT}$ transformations commute with the BRST transformations. Note that the transformations above are not to be identified with actual parity and time reversion. However, they are also such that ${\cal P}^2={\cal T}^2=1$, with an antilinear operator ${\cal T}$. From the discussion of the previous subsection, we have that the action $S_{RGZ}$ is Pseudohermitian
\cite{Mostafazadeh_2002_438}, i.e., $(S_{RGZ})^\dagger=\eta (S_{RGZ})\eta^{-1}$ for some Hermitian operator $\eta$. After the appropriate Legendre transformation, one could define a pseudohermitian Hamiltonian $H_{RGZ}$ operator. Being a nonhermitian operator, its corresponding evolution operator is evidently not unitary. However, according to the general theory of pseudohermitian Quantum Mechanics, unitarity may possibly be restored (at least in principle, for the case of real energy spectrum) with an appropriate redefinition of the inner product of the Hilbert space of the theory \cite{Mannheim:2009zj,Mostafazadeh_2010_07}, as discussed in Subsection \ref{sec:PTsymm-review}. For the case of complex spectrum, Pseudohermiticity implies that these eigenvalues come in complex-conjugate pairs, a fact which is easily seen from the RGZ propagators at tree level.

In the context of RGZ, the leading contribution to a metric operator can be easily calculated. The toy models of \cite{Mintz:2024soo} are simplifications of the quadratic RGZ action (\ref{eq:RGZ-quadratic}). With this in mind, it is not difficult to see that the positive operator
\begin{eqnarray}\label{eq:metric_RGZ-quad}
    \eta_{RGZ}^{quad}=\exp\left\{-2\theta\int d^{d-1}x
    \left[
    A_\mu^a(\Pi_V)_\mu^a-
    V_\mu^a(\Pi_A)_\mu^a
    \right]
    \right\},
\end{eqnarray}
where $\Pi_A$ and $\Pi_V$ are canonical momenta associated with the fields $A$ and $V$, and $\theta$ given by (\ref{eq:theta}), implements the intertwining relation $(S_{RGZ})^\dagger=\eta (S_{RGZ})\eta^{-1}$ at the quadratic level. Note that (\ref{eq:metric_RGZ-quad}) can be thought of as a sort of rotation in field space $(A,V)$ by an imaginary angle $2i\theta$. Then, we may use this metric operator to define Hermitian operators in ${\cal H}_\eta$, which are 
candidates for observables. A direct calculation (that closely generalizes those of \cite{Mintz:2024soo}) gives 
\begin{eqnarray}
\lambda_\mu^a&=&(\eta_{RGZ}^{quad})^{-1/2}\;A_\mu^a\;(\eta_{RGZ}^{quad})^{1/2}=A_\mu^a\cosh\theta + iV_\mu^a\sinh\theta,\nonumber\\
\eta_\mu^a&=&(\eta_{RGZ}^{quad})^{-1/2}\;V_\mu^a\;(\eta_{RGZ}^{quad})^{1/2}=-iA_\mu^a\sinh\theta + V_\mu^a\cosh\theta,
\end{eqnarray}
which coincide precisely with the quasiparticle operators
(\ref{eq:quasi-particles}). 

As an immediate consequence of the above construction, the quasiparticle operators $(\lambda_\mu^a)^\sharp=\lambda_\mu^a$
and $(\eta_\mu^a)^\sharp=\eta_\mu^a$. According
to Eq. (\ref{eq:spectral-function}), this implies that the
two-point functions $\langle\lambda_\mu^a\lambda_\nu^b\rangle$
and $\langle\eta_\mu^a\eta_\nu^b\rangle$ must have
nonnegative spectral functions. Indeed, we see that
this is the case, as can be seen from the propagators
(\ref{eq:propagators-quasiparticles}), as long as 
$\theta\in\mathbb{R}$ (or equivalently, when $(m^2-M^2)^2>8g^2N_c\gamma^4$).
 The important $\theta\not\in\mathbb{R}$ case (which comprises both GZ and the lattice-fitted RGZ) is much more subtle. Let us note that the colored fields $\lambda_\mu$ and $\eta_\mu$ are gauge dependent and may not be called observables in a strict sense, nor can their quantized excited states be directly identified with physical particles. 
 Furthermore, recall that actual observable fields in a gauge theory live in the BRST cohomology, to which we have no access in the original RGZ theory, since it does not have a BRST invariant action. In order to bridge this gap in the theory, we believe that a more consistent approach would be to use the BRST invariant formulation of the RGZ theory \cite{CapriEtAl2015_LCG,CapriEtAl2015_NilpotentBRST,CapriEtAl2016_MoreLCG,CapriEtAl2016_LocalBRST_Ah,CapriEtAl2016_A2min,CapriEtAl2017_RGZ_LCG,Capri:2017bfd,CapriEtAl2017_Nielsen,CapriEtAl2018_GaugeFixings,CapriEtAl2018_Universal,CapriEtAl2018_SYM_Stueckelberg,DudalEtAl2019_BRSTvacuum,CapriSorellaTerin2021_RGZ,DudalVercauteren2023_GapEq} to construct BRST invariant composite operators akin to glueball operators. However, such an argument would be more sophisticated and beyond the scope of the initial claims of this paper, which we regard as a preparation for this discussion. These further refinements will be discussed in a future publication.

%%%%%%%%%%%%%%%%%%%%%%
\section{Conclusions and perspectives}\label{sec:conclusions}

The RGZ Lagrangian can be considered as an effective theory for Yang-Mills theories which takes into account some relevant nonperturbative effects. In this paper, we reviewed the diagonalization of the RGZ Lagrangian in the quadratic approximation, which leads to the identification of quasiparticles of the theory. We showed how the longitudinal part of the Zwanziger auxiliary fields can play an important role not only in the Lorentz structure of the quasiparticle propagators, but also (and more importantly) in the diagonalization itself. Indeed, we proposed a version of the RGZ action in which the Zwanziger auxiliary fields are constrained to be transverse. We then showed that the quasiparticle operators are local in this theory, whereas this is not the case in the original RGZ theory (where the Zwanziger fields are not transverse). 

Such excitations can be properly interpreted, at least in the quadratic approximation, when there are no complex mass poles in the theory. This can be achieved by acknowledging that the action of the theory is not hermitian, but rather pseudohermitian. A crucial point is that we identified field operators ($\lambda_\mu$ and $\eta_\mu$) which have propagators with a non-negative spectral function, thus possessing a well-defined K\"all\'en-Lehmann representation. We interpret these properties as a strong indication that a proper continuation of RGZ to real time may be possible. With it, one could define a Hermitian Hamiltonian, a unitary evolution operator and Hermitian quasiparticle operators, as long as masses are real (at least in the deep UV regime, where the quadratic approximation could be acceptable). 

Of course, many details of the theory still have to be stated more precisely in order to take this discussion from an Euclidean path integral to a physical theory in real time. In particular, all of the Osterwalder-Schrader axioms are still to be checked explicitly in order to ensure the existence of a real-time theory with a well-defined Hilbert space. Also, the stability of these properties against perturbative corrections due to interactions must be shown. Finally, since RGZ is a gauge theory, the adequate Kugo-Ojima construction \cite{Kugo:1979gm} must also be properly outlined. In any case, we believe that the framework we sketched in the present work may possibly contribute to the discussion of bridging positivity violation in the infrared and asymptotically free states in the ultraviolet regime. An analogous study in the context of quadratic quantum gravity has been made in  \cite{kuntz2024unitarityptsymmetryquantum}.

Some interesting consequences of this work should be further explored. In particular, the quasiparticles can be used to construct scalar composite field operators which can be identified with glueball operators \cite{Sorella_i_particles,CapriEtAl2011Glueballs,bib:nele}. An obvious (but not trivial) direction to explore is that of the regions of parameter space in which mass poles are complex. In this case, the Hilbert space of the theory has to be constructed with care in order to avoid nonphysical behavior of would-be physical quantities. (For example, in the calculation of the glueball spectrum in \cite{Sorella_i_particles,DudalEtAl2010GlueballPropagators}, a complex, therefore nonphysical, contribution had to be excluded in a relatively {\it ad hoc} manner.) Complex mass poles should also be related to a (problematic\footnote{A nonhermitian metric operator $\eta\not=\eta^\dagger$ would be problematic. Indeed, such an operator may not be used to construct a sesquilinear $\eta-$inner product, rendering it unsuitable for a proper mathematical description of quantum systems.}) breaking of hermiticity of the metric operator $\eta$ in Eq. (\ref{eq:metric_RGZ-quad}), since the case of complex  masses also corresponds to $\theta\not\in\mathbb{R}$. Finally, the metric operator $\eta$ is typically not bounded, as is the case of (\ref{eq:metric_RGZ-quad}). This means that the Hilbert space ${\cal H}_\eta$ of the physical theory has to be defined accordingly. We believe that a proper study of all these matters is highly nontrivial, but necessary, as they will have important consequences not only for the study of quantum-mechanical ${\cal PT}$-symmetric systems, but also of confining Yang-Mills theories.

% --------------------------------------------------
\section*{Usage of AI}
% --------------------------------------------------

We have used Anthropic's Opus 5 model in order to search and organize a few references. We have personally checked each reference. For the present version of the manuscript, we also asked this model to make a pre-submission technical feedback. We partially implemented it, according to our agreement on each of the suggestions given.

% --------------------------------------------------
\section*{Acknowledgments}
% --------------------------------------------------

The authors thank Silvio Sorella, Letícia Palhares, and Marcelo Guimarães for clarifying discussions regarding the connection between i-particles and the glueball spectrum. This study was financed in part by the Coordenação de Aperfeiçoamento de Pessoal de Nível Superior – Brasil (CAPES) – Finance Code 001.

\bibliography{biblio}

\end{document}